# Emergent Coherence at the Edge of Magnetism: Low-Doped La$_{2-x}$Sr$_x$CuO$_{4+\delta}$ Revisited

E.Yu. Beliayev[1,2*], Y.K. Mishra[2], I.A. Chichibaba[3], I.G. Mirzoiev[1], V.A. Horielyi[1], A.V. Terekhov[1]

[1]B. Verkin Institute for Low Temperature Physics and Engineering of the Nat. Acad. of Sciences of Ukraine.

[2]Smart Materials Group, Mads Clausen Institute, University of Southern Denmark, Sønderborg.

[3]National Technical University «Kharkiv Polytechnic Institute», Kharkiv, Ukraine.

e-mail*: *beliayev@ilt.kharkov.ua*

## Abstract

The La$_{2-x}$Sr$_x$CuO$_{4+\delta}$ (LSCO) system provides a unique experimental setting for exploring how magnetism, superconductivity, and disorder jointly shape charge transport in a doped Mott insulator. Transport measurements in lightly doped and oxygen-enriched LSCO reveal a strongly insulating normal state governed by variable-range hopping, accompanied by pronounced nonlinear current–voltage characteristics and, at low temperatures, current-induced negative differential resistance. With increasing carrier concentration, these features evolve into regimes characterized by granular and percolative superconductivity near the threshold of bulk superconductivity and, eventually, into a homogeneous strange-metal state close to optimal doping.

Throughout this evolution, the transport response shows marked sensitivity to disorder, electronic inhomogeneity, and external control parameters, such as bias current and magnetic field. Rather than reflecting a sequence of sharply distinct phases, the observed transport regimes form a continuous crossover from a localization-dominated insulating state to granular superconductivity and further to a coherent metallic state. This crossover is driven primarily by the progressive enhancement of electronic screening, inter-region coupling, and superconducting connectivity, rather than by abrupt changes in the underlying microscopic scattering mechanisms.

Taken together, the available transport data provide a coherent experimental basis for understanding how disorder and mesoscale electronic inhomogeneity organize charge transport and superconductivity across the LSCO phase diagram, underscoring the central role of percolation and nonequilibrium effects in underdoped cuprates.

## Introduction

Among the family of layered cuprate superconductors, La$_{2-x}$Sr$_x$CuO$_{4+\delta}$ (LSCO) is used as a prototypical system for studying the interplay between antiferromagnetism, superconductivity, and disorder. In its undoped stoichiometric form ($x = 0$, $\delta = 0$), La$_2$CuO$_4$ is a charge-transfer antiferromagnetic insulator (often described within the Mott–Hubbard paradigm) with a Néel temperature $T_N \approx 325$ K and a charge-transfer gap in the electronic spectrum [1]. Its low-temperature orthorhombic crystal structure and anisotropic antiferromagnetic order, including weak ferromagnetic canting driven by Dzyaloshinskii–Moriya interactions, have been characterized in detail in high magnetic fields [2]. Light Sr doping (substituting Sr for La) introduces mobile holes into the CuO$_2$ planes, which rapidly suppresses the long-range AFM order and, upon reaching a critical carrier concentration, gives rise to high-temperature superconductivity. Excess oxygen intercalation offers an alternative doping route, as oxygen intercalation in $La_{2-x}Sr_xCuO_{4+\delta}$ leads to genuine chemical phase separation driven by oxygen diffusion, producing superconducting and antiferromagnetic regions on mesoscopic length scales [3]. By contrast, in Sr-doped cuprates static chemical segregation is absent, while spatially modulated electronic states such as charge and spin stripes, stabilized or pinned by lattice distortions, have been directly observed by neutron scattering [4]. On the electronic side, Mohottala *et al.* [5] provided clear evidence for electronic phase separation in superoxygenated LSCO: for the Sr contents studied, adding excess oxygen ($\delta$) raised the superconducting $T_c$ to ~40 K while simultaneously inducing a static spin-density-wave (SDW) order that onsets near 40 K. Notably, interstitial oxygen can self-organize, promoting oxygen-rich and oxygen-poor regions. If cooled slowly, $La_2CuO_{4+\delta}$ undergoes phase separation for oxygen contents in the range $\delta \simeq 0.01–0.06$ into two phases: one nearly stoichiometric, antiferromagnetic phase, and one O-rich superconducting [3].

Experimentally, the hole concentration $p$ in LSCO is most directly controlled by Sr substitution ($p \approx x$), while for oxygen-doped crystals ($La_2CuO_{4+\delta}$), estimating $p$ is more subtle. Chen *et al.* [1] established an empirical relation $p \approx 2\delta$, based on Hall-effect measurements, which is often used as an approximation at small $\delta$. At higher $\delta \gtrsim 0.03$, however, the situation becomes more complex: interstitial oxygen atoms no longer contribute two holes per atom due to site-dependent doping efficiency and increasing electronic inhomogeneity. Li et al. [6] demonstrated that only ~1.3 holes per oxygen atom are introduced beyond a critical concentration of $p_s \approx 0.06$, at which enhanced charge clustering and electronic inhomogeneity become important. As a result, $T_N$, the AFM transition temperature, is often used as a practical proxy for $p$, enabling comparisons between Sr- and O-doped LSCO along a unified hole concentration axis [7].

The standard phase diagram of LSCO encompasses an AFM phase at very low hole doping ($p$) up to $p \approx 0.02$ (approximately $x \lesssim 0.02–0.03$) and a dome-shaped superconducting (SC) region for $0.05 \lesssim x \lesssim 0.30$, with a maximum transition temperature $T_c^{max} \sim 38$ K around $x \approx 0.15$ [7]. On the underdoped side, a pseudogap regime emerges below a characteristic temperature $T^*$ above $T_c$. Between the AFM and SC phases, for $x \approx 0.02–0.05$, the system enters a spin-glass or cluster AFM state: a disordered magnetic phase in which static short-range magnetic order persists and spins freeze locally at low temperatures (typically $T_f \sim 10–15$ K) [8]. Classic studies, such as those by Takagi et al. [9], complemented by Hall-effect studies by Ando et al. [10], established the broad trends in LSCO's transport across this phase diagram: the resistivity rises steeply at low doping (insulating behavior) and evolves to an approximately linear-in-$T$ dependence near optimal doping, - a behavior widely regarded as the hallmark of the "strange metal" state, characterized by the breakdown of conventional Fermi-liquid theory and the dominance of unconventional scattering processes.

At very low hole concentrations ($p \lesssim 0.02$), realized in LSCO at Sr contents $x \lesssim 0.02$ or under extremely weak oxygen doping, transport measurements reveal a strongly localized insulating state. The resistivity $\rho(T)$ increases upon cooling and is often described phenomenologically in terms of Mott-type variable-range hopping (VRH) behavior $\rho(T) \propto exp[(T_0/T)^{1/4}]$ at sufficiently low temperatures [9]. This insulating behavior is commonly attributed to carrier localization in a disordered and magnetically correlated background, arising from the combined effects of ionic disorder and the antiferromagnetic (AFM) potential. Notably, Hall-effect measurements reveal a distinct anomaly in the Hall coefficient at the Néel temperature $T_N$, signaling a reconstruction of the electronic response associated with the onset of long-range AFM order [10]. Complementary resistivity and magnetotransport studies reveal that the temperature dependence of $\rho(T)$ changes its character across $T_N$: upon warming through $T_N$, the resistivity becomes less insulating, indicating enhanced carrier mobility once the long-range AFM order is suppressed [11]. In this sense, the establishment of AFM order tends to localize charge carriers, whereas thermal disordering of the AFM state (for $T > T_N$) partially relieves this localization. This close correlation between magnetic ordering and transport behavior underscores the strong interplay between spin and charge degrees of freedom in lightly doped LSCO.

Even within the insulating regime, anomalies emerge at very low temperatures in disordered samples. In some oxygen-enriched LSCO crystals with extremely low average hole doping ($p \approx$ 0.002–0.004), deviations from standard VRH behavior appear below ~20–25 K [12] [13]. The resistivity $\rho(T)$ develops a downward curvature upon cooling, deviating from the Mott law and instead showing a more rapid, activated drop. In one case, a finite activation energy $\Delta \approx 30$ K was extracted, suggesting the opening of a mini-gap [12] [13]. This behavior was interpreted as potentially reflecting the development of superconducting correlations in an inhomogeneous background. While the samples remain globally insulating and do not reach zero resistance, the low-$T$ deviation from the VRH form may indicate the emergence of an additional conduction channel or gap-like feature at low energies.

Moreover, pronounced nonlinear transport effects accompany this regime. Current–voltage measurements at low temperatures reveal a transition to negative differential resistance (NDR) in the same low-doped, oxygen-rich LSCO crystals [13]. At currents above a threshold value, the voltage decreases with increasing current, indicating a sharp change in the differential conductance. This behavior is reproducible within a limited range of temperatures and doping levels and is consistent with a nonequilibrium electronic response of an inhomogeneous conducting network. However, a careful analysis is required to distinguish intrinsic transport effects from possible self-heating contributions, as discussed in [13].

Structural disorder can further enhance phase inhomogeneity, even when chemical dopants remain macroscopically well mixed. Local-probe studies, particularly $^{63}$Cu NQR measurements, have provided direct evidence for pronounced spatial variations in the local hole concentration on nanometer length scales in Sr-doped LSCO [14]. This indicates that a substantial part of the inhomogeneity is electronic in origin and may persist even in the absence of macroscopic chemical dopant segregation. Such electronically driven variations in hole density create regions that are locally more or less doped, predisposing the system to micro-phase separation in the electronic subsystem. Even before global superconductivity is established, hole-rich regions can support local superconducting pair correlations, while surrounding areas remain insulating or magnetically ordered. Transport signatures such as partial resistivity drops, activation-like behavior, and nonlinear current–voltage characteristics are consistent with the presence of short-lived or spatially confined superconducting correlations in these nanoscale regions, which effectively act as microscopic superconducting inclusions embedded in an antiferromagnetic insulating background.

As Sr doping increases toward a critical level, $x_c \approx 0.05$ ($p_c \approx 0.05$ holes/Cu), LSCO undergoes a transition from the insulating spin-glass regime into a state with global superconductivity. In real samples, particularly polycrystalline and oxygen-doped crystals, this superconducting percolation threshold is smeared and broadened by disorder. For example, ceramic LSCO samples near $x_c$ often display a two-step resistive transition: the resistivity begins to drop at an onset temperature $T_c^{onset}$, when local superconductivity emerges in isolated regions or grains, but it does not reach zero until a lower temperature $T_c$, at which a continuous superconducting path percolates through the sample. The initial drop reflects the formation of isolated superconducting clusters, while the final transition signals the establishment of long-range phase coherence through Josephson coupling between them. This behavior is characteristic of a granular superconductor, where macroscopic superconductivity is governed by weak inter-cluster links and the establishment of long-range phase coherence [15]. It should also be noted that the apparent threshold $x_c$ is not a universal constant: in practice, it can shift and broaden depending on crystal quality, residual oxygen nonstoichiometry/ordering, and thermal history, which all modify the effective disorder level and percolation of superconducting regions.

Further support for this granular scenario comes from magnetotransport studies. In strongly inhomogeneous LSCO near threshold doping, the magnetotransport in lightly doped LSCO often exhibits history-dependent magnetoresistance with pronounced memory effects, reflecting glassy charge dynamics in a strongly inhomogeneous insulating state [16]. Such phenomenology is broadly consistent with a granular/weak-link scenario in which global phase coherence emerges only after inter-island coupling becomes effective [15]. Similar nonlinear magnetotransport phenomena, arising in spatially inhomogeneous superconductors and often associated with percolative connectivity of superconducting regions, have been widely reported in 2D superconductors [17], as well as in other granular or disordered systems [18], supporting the view that the superconducting transition in disordered LSCO is governed by percolation effects rather than by the uniform establishment of global superconducting coherence.

At optimal doping (around $x \approx 0.15$), LSCO's normal-state resistivity exhibits the well-known "strange metal" behavior: $\rho(T)$ is approximately linear in temperature over a broad range, extending from just above $T_c$ up to room temperature and beyond [9] [19]. This robust linear-in-$T$ dependence is observed not only in high-quality single crystals but also in polycrystalline and moderately disordered samples, indicating that it cannot be trivially attributed to sample imperfections. More generally, these observations indicate that non-Fermi-liquid transport is not an inevitable consequence of strong electronic correlations alone. Depending on the degree of disorder, effective dimensionality, and the evolution of the low-energy electronic structure, cuprates may exhibit either strange-metal behavior with a linear temperature dependence of the resistivity [19] or a conventional Fermi-liquid-like response with $\rho \propto T^2$, in the normal state [20]. Upon approaching the superconducting transition from above, within a narrow temperature window above $T_c$, even optimally doped LSCO can exhibit deviations from strict linearity in $\rho(T)$, along with enhanced sensitivity to applied currents and magnetic fields [21]. Such features are commonly attributed to superconducting fluctuations and phase-incoherent pairing above $T_c$ [22] [23], and may be further influenced by residual electronic inhomogeneity [24] [25]. In particular, a slight downward curvature of $\rho(T)$ just above $T_c$, reflecting the onset of paraconductivity due to short-lived Cooper pairs, has been widely reported in cuprates and is consistent with fluctuation-driven transport in the vicinity of the superconducting transition, as described by the Aslamazov–Larkin [26] and Maki–Thompson [27] [28] fluctuation conductivity frameworks and observed experimentally in LSCO and related cuprates [29] [30].

Despite decades of intensive research and a fairly well-mapped global phase diagram for LSCO [7], many fundamental questions remain regarding the microscopic coexistence and evolution of local phases, especially in disordered environments. How do antiferromagnetic, superconducting, and metallic regions coexist or compete on the nanoscale? What governs the onset of global coherence in a landscape that is granular or phase-separated? Can spatially non-uniform carrier distributions stabilize metastable or "hidden" phases that are absent in the clean limit? These issues gain special significance in light of experimental evidence for phase-separated and spatially inhomogeneous states [5] that do not fit neatly into the canonical phase diagram and suggest local violations of the global phase rules.

In this review, we synthesize experimental results obtained from extensive studies performed at the B. Verkin Institute for Low Temperature Physics and Engineering (ILTPE, Kharkiv) on LSCO single crystals and ceramics, focusing on magnetism and electronic transport in low- and near-optimally doped LSCO+δ. These studies encompass oxygen-doped $La_2CuO_{4+\delta}$ and $La_{2-x}Sr_xCuO_{4+\delta}$ single crystals, where mobile oxygen defects promote electronic and magnetic phase inhomogeneity [11] [12] [13] [31] [32], as well as Sr-doped LSCO ceramics, in which quenched compositional disorder and microstructural granularity govern superconducting and transport properties across a wide doping range [33] [34] [35] [36] [37]. Special emphasis is placed on phase inhomogeneity, nonlinear transport phenomena, granular superconductivity, and the emergence of mixed or metastable states at the boundary between antiferromagnetism and superconductivity.

Within this broader context, it is important to note that, in the clean-limit idealization of a spatially homogeneous system, several boundaries in the LSCO phase diagram are often discussed in terms of a putative zero-temperature quantum phase transition. In real LSCO, however, quenched disorder and intrinsic inhomogeneity tend to smear these boundaries, so that transport and superconducting coherence evolve primarily via continuous crossovers and percolation-driven connectivity changes.

The paper is organized around the evolution of electronic transport and superconductivity in LSCO as a function of hole doping, progressing from the antiferromagnetic insulating regime at $x \to 0$ to the optimally doped superconducting state near $x \approx 0.15$. At low doping, we focus on transport dominated by strong localization, pronounced nonlinearity, and current-induced negative differential resistance, reflecting the highly inhomogeneous and nonequilibrium character of the system. In the vicinity of the superconducting threshold, the emphasis shifts to the emergence of superconductivity in a granular and percolative form, shaped by electronic inhomogeneity and disorder. Upon further doping toward optimal composition, transport crosses over into a coherent strange-metal regime with linear-in-temperature resistivity, highlighting the evolving roles of disorder, electronic screening, and inter-region coupling across the LSCO phase diagram. Within each doping regime, we review the corresponding magnetic properties, the development of superconducting regions, the nature of the superconducting transition in the presence of disorder, and the normal-state transport behavior, and we discuss how disorder qualitatively modifies the magnetic, superconducting, and transport properties across the phase diagram. The Discussion section synthesizes these results within a unified phenomenological framework, while the Conclusion summarizes the main physical insights.

## Doping-Dependent Disorder, Granularity, and Charge Transport in $La_{2-x}Sr_xCuO_{4+\delta}$

### Low-Doping Regime ($La_{2-x}Sr_xCuO_4$ and $La_2CuO_{4+\delta}$)

At very low hole doping level (well below $p \approx 0.08$ per Cu), $La_{2-x}Sr_xCuO_4$ (LSCO) resides in a strongly underdoped regime in which disorder and intrinsic inhomogeneity, together with the antiferromagnetic background, strongly affect transport and superconductivity [7]. In this regime, the in-plane resistivity becomes strongly insulating at low temperatures and is commonly described in terms of carrier localization and hopping-type transport [7] [9]. In particular, for lightly Sr-doped LSCO samples with $x \approx 0.03$–$0.05$, the low-$T$ resistivity can often be described by the three-dimensional Mott variable-range hopping law, $\rho(T) \propto exp[(T_0/T)^{1/4}]$, below ~20–30 K, as representative behavior of the lightly doped regime [7].

Typical temperature dependences of the resistivity in lightly Sr-doped LSCO indeed exhibit Mott variable-range hopping behavior over a broad temperature range, with systematic deviations emerging at lower temperatures and under elevated bias currents (Fig. 1). These deviations foreshadow the onset of strongly nonlinear and nonequilibrium transport behavior discussed below.

All measurements were performed on bulk ceramic LSCO samples with typical dimensions of $11 \times 3 \times 4$ mm$^3$, using a four-probe configuration and a Keithley nanovoltmeter, allowing reliable measurements over a wide range of bias currents without significant self-heating.

Reported VRH parameters indicate very strong localization: characteristic temperatures $T_0$ on the order of $10^3$–$10^4$ K and localization lengths of only a few angstroms (e.g., $L_c \approx 0.4$–$0.8$ nm) in lightly doped $La_2CuO_4$-based systems [13] [32]. In such insulating antiferromagnetic or spin-glass regimes, applying a magnetic field often reduces the resistivity, resulting in a pronounced negative magnetoresistance that is commonly attributed to magnetic-field-induced suppression of quantum interference and spin-dependent effects in hopping transport [38] [39] [40], with analogous magnetic-field suppression of interference effects also observed in disordered metallic systems [41]. For example, Dalakova et al. [32] reported a quadratic negative magnetoresistance up to 1.75 T in antiferromagnetic $La_2CuO_{4+\delta}$ (LCO+δ) crystals ($T_N \approx 265$ K) over the temperature range 18–100 K, which they interpreted in terms of field-induced alignment of localized electron spins and the resulting suppression of spin-interference effects in the VRH regime. By contrast, in lightly Sr-doped LSCO ($x \approx 0.01$–0.03), a sizable negative in-plane MR develops once Néel order is established; high-field measurements ($H \parallel c$) demonstrate that reorientation of weak-ferromagnetic canting and associated domain structures leads to a large, saturating negative magnetoresistance at fields of several tens of tesla, with low-field hysteresis reflecting domain reconfiguration [42]. In the spin-glass regime ($x \approx 0.03$–0.05), the magnetoresistance remains predominantly negative and tends to saturate by ~40 T; even moderate laboratory fields (e.g., 14 T) can already reduce ρ by ~20–40% at low T, reflecting the dominant role of spin-related scattering in this disordered magnetic state [43]. Overall, the emergence of strong negative magnetoresistance across these lightly doped regimes highlights the intimate coupling between charge transport, magnetic correlations, and disorder in LSCO and related cuprates [7].

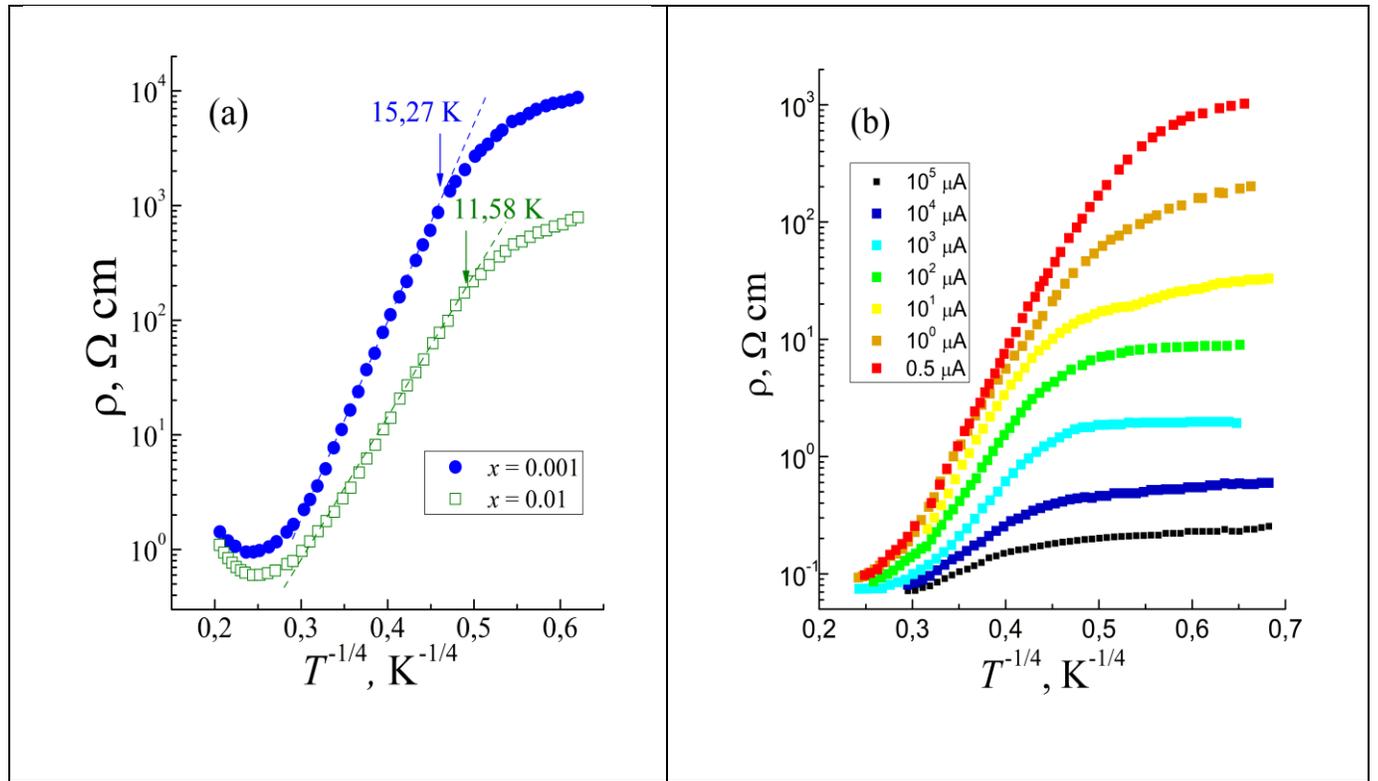

Fig. 1. Temperature dependence of the resistivity ρ(T) for lightly Sr-doped $La_{2-x}Sr_xCuO_4$ samples with $x = 0.001$ and $x = 0.01$, shown in Mott variable-range hopping coordinates. Straight lines indicate the temperature range where 3D Mott VRH behavior holds. The right panel illustrates the evolution of ρ(T) for $x = 0.01$ at different measurement currents, highlighting deviations from the VRH regime induced by the currents from 0.5 (upper curve) to 100 000 μA (lower curve). Data replotted from Ref. [37]

**Granularity and Two-Stage Superconducting Transition**

*General concept: electronic phase separation and granular superconductivity*

A characteristic feature of the underdoped regime in La-based cuprates is the emergence of spatially inhomogeneous superconductivity, which develops well before a coherent bulk superconducting state is established. Owing to chemical disorder and electronic phase separation, superconducting (SC) clusters or islands are believed to form within an insulating antiferromagnetic matrix even when the nominal hole concentration remains below the percolation threshold for global superconductivity, consistent with μSR evidence for microscopic phase separation in lightly doped LSCO [8]. Experiments on both Sr- and O-doped $La_2CuO_4$ further indicate that doped holes are distributed inhomogeneously on mesoscopic length scales, with pronounced spatial variations of the local hole concentration revealed by NQR measurements [14], producing a heterogeneous electronic landscape in which locally superconducting regions can coexist with magnetically ordered insulating domains. In such a system, superconductivity is intrinsically granular, and global phase coherence can only emerge through Josephson coupling and percolation between superconducting regions. Here, we use the term granular superconductivity in the sense of intrinsic electronic granularity: superconducting puddles or clusters emerge from nanoscale/mesoscale variations of local hole density and competing orders, rather than from extrinsic morphological granularity of a ceramic microstructure or imperfect electrical contacts.

It is instructive to contrast intrinsic electronic granularity in underdoped cuprates with extrinsic granularity arising from microstructural inhomogeneity in polycrystalline samples. In the $La_{1.85}Sr_{0.15}CuO_4$ ceramic sample, Dalakova et al. [36] demonstrated a pronounced two-stage superconducting transition, where the onset of superconductivity within individual grains precedes the establishment of global phase coherence mediated by intergranular Josephson coupling. The resistive transition is strongly broadened and exhibits a clear separation between the intragranular superconducting temperature $T_{c0}$ and a lower temperature $T_{cJ}$ associated with percolative Josephson paths. While phenomenologically similar two-step transitions are also observed in electronically inhomogeneous underdoped LSCO and oxygen-doped $La_2CuO_{4+\delta}$, their microscopic origins are fundamentally different: in ceramic samples, granularity is structural and extrinsic, whereas in single crystals it emerges from intrinsic electronic phase separation.

In this context, it is useful to distinguish between two qualitatively different routes to inhomogeneity in La–based cuprates. In Sr-doped LSCO, the immobility of the dopant on experimental time scales precludes true chemical phase separation; instead, disorder and competing electronic correlations promote electronic phase separation, which is manifested as nanoscale variations in the local hole density and the coexistence of superconducting and magnetic regions. In contrast, excess oxygen doping enables chemical phase separation, in which mobile interstitial oxygen atoms self-organize into regions with distinct structural and electronic properties, characterized by oxygen-rich and oxygen-poor domains. While both routes ultimately lead to granular superconductivity in the underdoped regime, the underlying mechanisms and the resulting depth of phase separation are fundamentally different.

*Oxygen-doped* $La_2CuO_{4+\delta}$: *deep chemical phase separation*

This phase separation is particularly pronounced in $La_2CuO_{4+\delta}$, where interstitial oxygen is highly mobile and can aggregate into oxygen-rich regions [3] [44]. Neutron diffraction studies have demonstrated that upon cooling, $La_2CuO_{4+\delta}$ crystals spontaneously separate into two distinct phases: one phase is essentially undoped $La_2CuO_4$ with long-range antiferromagnetic order, while the other is an oxygen-rich phase that becomes superconducting below $T_c \approx 38–40$ K [44]. Complementary NQR measurements [14] further reveal pronounced spatial variations of the local hole concentration in these materials, indicating strong electronic inhomogeneity on mesoscopic length scales. Remarkably, this occurs even in samples with very low Sr content. As a result, even $La_{2-x}Sr_xCuO_{4+\delta}$ samples with very low Sr concentration can exhibit a superconducting transition near 40 K once a small amount of excess oxygen is introduced, because the local hole concentration within the oxygen-rich regions approaches optimal doping despite the much lower average carrier density.

Direct evidence for this phase-separated state was provided by Mohottala et al. [5], who showed that in superoxygenated LSCO the superconducting regions consistently exhibit $T_c \approx 40$ K, while the coexisting magnetic phase displays spin-density-wave order at the same temperature. This behavior is consistent with a nearly binary system composed of optimally doped superconducting regions and magnetically ordered regions, with an effective doping near $p \approx 1/8$, where the relative volume fractions are controlled by oxygen ordering and thermal history. The superconducting volume fraction is sensitive to thermal history, underscoring the importance of oxygen mobility and ordering kinetics [44], while the emergence of global superconductivity proceeds via percolative connectivity of superconducting regions [5]. Such oxygen-doped systems can therefore be viewed as granular superconductors composed of superconducting puddles embedded in an insulating background, weakly coupled via Josephson links [13].

Sr-*doped* $La_{2-x}Sr_xCuO_4$: *disorder-induced electronic granularity*

A similar, though less extreme, form of granularity also arises in Sr-doped LSCO near the superconductor–insulator threshold. In this case, the Sr dopants are immobile and do not produce true chemical phase separation; instead, electronic inhomogeneity emerges from quenched disorder and competing magnetic correlations. For underdoped LSCO with $x \approx 0.05$–0.06, superconductivity typically onsets at $T_c^{onset} \approx 15$–20 K, while zero resistance is reached only at much lower temperatures ($T_{c0} \approx 5$–7 K for $x \approx 0.06$). The large separation between $T_c^{onset}$ and $T_{c0}$ reflects the fact that only a limited fraction of the sample becomes superconducting at higher temperatures, whereas global superconductivity requires percolation and phase locking of these regions at lower $T$.

Consistent with this picture, μSR measurements reveal an incomplete superconducting volume fraction in lightly doped samples, with static magnetic correlations persisting well above $T_{c0}$ [8]. These results point to a "patchy" superconducting state, where superconducting regions coexist with antiferromagnetic or spin-glass domains, despite the absence of macroscopic chemical phase separation.

It is worth noting, however, that while Sr dopants are immobile on experimental time scales, their spatial distribution may not be purely random. During crystal growth and high-temperature annealing, Sr ions are mobile and may partially self-organize under the influence of electronic and lattice correlations, potentially favoring local hole concentrations close to commensurate values (e.g., $p \approx 1/8$). Upon cooling, such an inhomogeneous dopant configuration becomes quenched into the antiferromagnetic $La_2CuO_4$ matrix. In this sense, Sr substitution may promote an electronically driven *precursor* to phase separation, which can later be amplified by excess oxygen doping, leading to a deeper, chemically stabilized phase separation.

*Transport manifestation*: *two-stage transition and broad* $\Delta T$

One immediate consequence of this granularity in electron transport is a broad, two-step resistive transition as the temperature is lowered into the superconducting state. In oxygen-doped $La_2CuO_{4+\delta}$ and $La_{2-x}Sr_xCuO_{4+\delta}$ with low average doping, resistivity measurements frequently show a partial drop in $\rho(T)$ tens of kelvins above the eventual zero-resistance transition, reflecting the onset of superconductivity in isolated regions that do not yet form a macroscopic current path [13]. Upon further cooling, the resistivity decreases more gradually and vanishes only at lower temperatures, when global phase coherence is established through percolation of superconducting clusters.

In practical terms, the granular nature of superconductivity in the underdoped regime leads to unusually broad and nonuniform superconducting transitions, in stark contrast to optimally doped cuprates, where transition widths $\Delta T$ are typically below 1 K. Oxygen-doped $La_{2-x}Sr_xCuO_{4+\delta}$ provides a particularly striking example: although the superconducting onset occurs near 30–40 K, bulk superconductivity develops only at substantially lower temperatures, reflecting a strongly inhomogeneous superconducting state [5]. By contrast, Sr substitution tends to produce a more uniform (though still disordered) carrier distribution, so Sr-doped LSCO generally exhibits narrower transitions at comparable average hole concentrations. Direct evidence for the percolative nature of superconductivity in oxygen-doped La-based cuprates has been provided by surface resistivity measurements on $La_2CuO_{4.06}$, which reveal a characteristic two-step transition with distinct critical temperatures associated with intra-puddle and inter-puddle superconducting coherence, consistent with successive percolation thresholds in a phase-separated system ($T_{c1} \approx 27$ K and $T_{c2} \approx 13$ K) [45].

A similar two-stage superconducting transition is also observed in ceramic $La_{2-x}Sr_xCuO_4$ samples with $x =$ 0.05–0.15, where granularity is of extrinsic, microstructural origin. In such systems, superconductivity first develops within individual grains at $T_{c0}$, while global zero resistance is established only at a lower temperature, $T_{cJ}$, via percolative Josephson coupling between grains [36]. Reducing the Sr concentration enhances compositional and structural disorder, leading to a strong broadening of the resistive transition and a pronounced sensitivity to transport current and magnetic field. While phenomenologically similar to intrinsic granularity in underdoped single crystals, this behavior reflects a fundamentally different physical mechanism.

### Non-Linear Transport and NDR Phenomenon

A pronounced manifestation of electronic and magnetic inhomogeneity in the lightly doped regime of La-based cuprates is the emergence of non-Ohmic transport and current-driven negative differential resistance (NDR) at low temperatures. Such behavior has been systematically reported both in oxygen-doped $La_2CuO_{4+\delta}$ (LCO+δ) [11] [12] [13] as well as in very lightly Sr-doped $La_{2-x}Sr_xCuO_4$ (LSCO) [35] [37], albeit with important differences in the underlying physical mechanisms. In LCO+δ, NDR is commonly associated with current-induced redistribution of oxygen-rich superconducting regions and the formation of highly conductive filaments, whereas in lightly Sr-doped LSCO, it is generally attributed to nonequilibrium transport and self-heating effects in a strongly localized, percolative electronic system. Representative current–voltage characteristics demonstrating current-driven negative differential resistance in lightly Sr-doped LSCO are shown in Fig. 2 (data replotted from Ref. [37]). For samples with $x = 0.001$ and 0.01, pronounced deviations from Ohmic behavior and the emergence of NDR are observed at low temperatures ($T = 4.4$ and 10 K), above a characteristic threshold current density $J_c$.

At $T = 30$ K, the current–voltage characteristics exhibit a pronounced current-induced change of slope at high currents. For $x = 0.01$, the CVC remains monotonic, showing a clear deviation from the low-current Ohmic regime ($U \propto J$) without developing a maximum. By contrast, for $x = 0.001$, the CVC shows a weak local maximum in $U(J)$, followed by a slight decrease in voltage with increasing current. Although this decrease amounts to only a factor of a few on a logarithmic scale, it represents an incipient negative differential resistance. Importantly, the NDR-like feature emerges only after an extended regime of positive differential resistance, where $U(J)$ is approximately linear on a *log–log* plot over nearly three decades of current. This behavior indicates that the negative slope at $T = 30$ K is weak and much weaker than the well-developed NDR observed at lower temperatures.

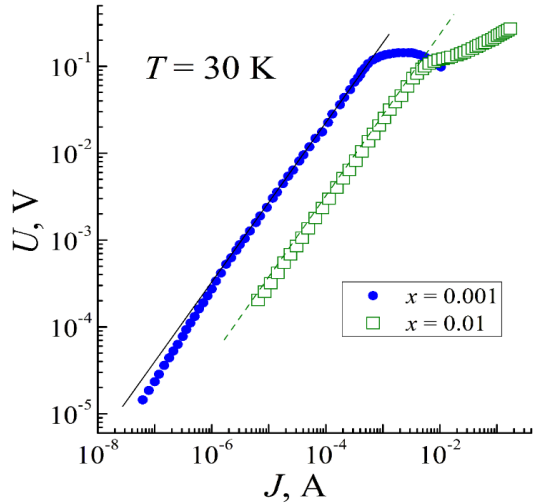

Fig. 2. Current-voltage characteristics of La$_{2-x}$Sr$_x$CuO$_4$ samples with $x = 0.001$ and $x = 0.01$ measured at $T = 30$ K. The characteristics exhibit pronounced nonlinear behavior and a current-induced change of slope at high currents. Straight lines indicate the low-current Ohmic regime. Data replotted based on Ref. [37].

An important aspect of these data is their pronounced temperature dependence. At elevated temperatures ($T = 30$ K), the current–voltage characteristics exhibit strong nonlinearity and a current-induced change of slope at high bias currents, but no well-developed negative differential resistance is observed (Fig. 2). By contrast, upon lowering the temperature the $I$–$V$ characteristics undergo a qualitative change: at $T = 4.4$ K and 10 K a clear current-driven negative differential resistance develops above a characteristic threshold current density $J_c$ (Fig. 3). This comparison demonstrates that the NDR phenomenon in lightly Sr-doped LSCO is intrinsically a low-temperature effect.

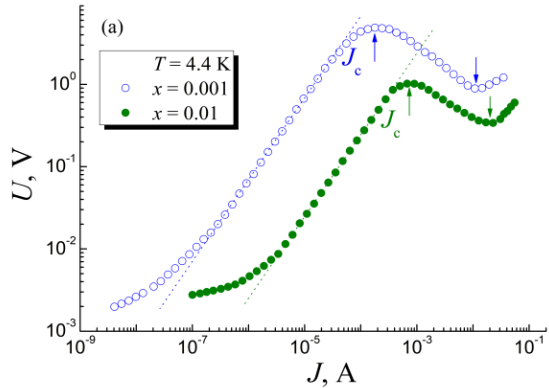 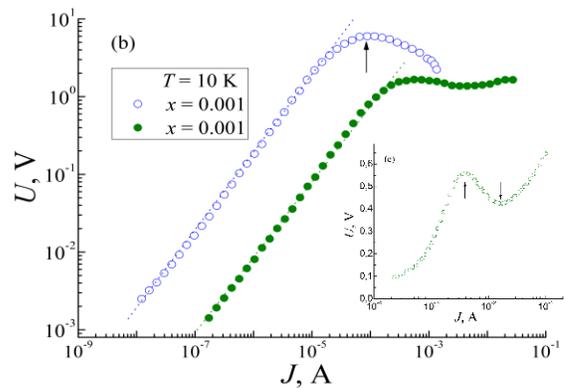

Fig. 3. Current–voltage characteristics of lightly Sr-doped La$_{2-x}$Sr$_x$CuO$_4$ samples with $x = 0.001$ and $0.01$ measured at low temperatures $T = 4.4$ K and 10 K. At sufficiently high currents, a pronounced current-driven negative differential resistance develops above a characteristic threshold current density $J_c$, as evidenced by the appearance of a maximum in $U(J)$ followed by a decrease of voltage with increasing current. Straight lines indicate the low-current Ohmic regime. The inset illustrates the evolution of the $I$-$V$ characteristics for $x = 0.01$ in semi-logarithmic representation. Data replotted based on Ref. [37].

This comparison demonstrates that the emergence of NDR in lightly Sr-doped LSCO is a distinctly low-temperature phenomenon and cannot be attributed to trivial contact effects or geometric nonlinearities, since the same samples and contact configurations exhibit no developed NDR at higher temperatures, despite the persistence of non-Ohmic transport. The disappearance of NDR at higher temperatures indicates that the low-$T$ NDR reflects a nonequilibrium electronic response of a strongly disordered antiferromagnetic system driven far from equilibrium by the applied current. In this context, a closely related electric-field–driven crossover between strong and weak electron localization, accompanied by pronounced nonlinearity of the current–voltage characteristics and a change in the sign of magnetoresistance, was previously demonstrated in percolating gold films near the percolation threshold, where the applied electric field effectively suppresses tunneling barriers between metallic clusters [46]. Consistent with this picture, additional insight into the nature of the low-temperature nonlinear transport in lightly Sr-doped LSCO is provided by magnetoresistance measurements performed at different bias currents. As shown in Fig. 4, the magnetoresistance at $T = 4.4$ K exhibits a pronounced dependence on the transport current for both $x = 0.001$ and $x = 0.01$.

Notably, in Sr-doped LSCO, this behavior develops in the absence of macroscopic chemical phase separation detectable on structural or compositional length scales, suggesting a physical mechanism distinct from that operative in oxygen-doped $La_2CuO_{4+\delta}$ systems, discussed below. With increasing current, the magnetoresistance is progressively suppressed, and, in the most-weakly doped sample ($x = 0.001$), it even changes sign from positive to negative.

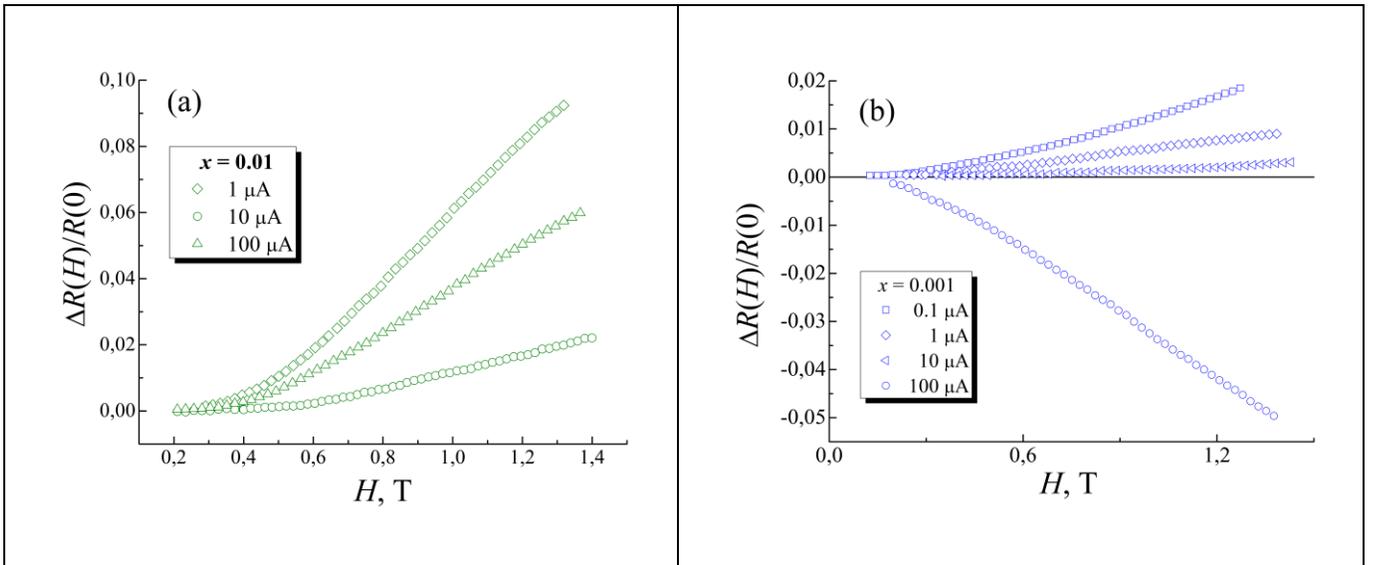

Fig. 4. Magnetoresistance of $La_{2-x}Sr_xCuO_4$ samples with $x = 0.001$ and 0.01 measured at $T = 4.4$ K for different values of the transport current in the configuration $J \parallel H$. A strong current dependence of the magnetoresistance is observed in both samples; however, the suppression and sign reversal of the magnetoresistance at high bias currents occur only in the lowest-doped sample ($x = 0.001$), highlighting the importance of current-driven nonequilibrium transport effects. Data replotted based on Ref. [37].

Such a strong current dependence of the magnetoresistance demonstrates that the electronic transport in lightly Sr-doped LSCO at low temperatures is strongly governed by current-driven, nonstationary transport effects. The suppression and eventual sign reversal of the magnetoresistance with increasing bias current indicate a current-driven modification of the effective transport regime in a disordered antiferromagnetic background. This behavior is consistent with scenarios involving electron overheating, current-induced hot-spot formation, and percolative hopping conduction in a disordered antiferromagnetic background, rather than with mechanisms dominated by the activation or suppression of superconducting links. These features further distinguish the nonlinear transport in Sr-doped LSCO from that observed in oxygen-doped $La_2CuO_{4+\delta}$, where macroscopic phase separation and superconducting inclusions play a central role, as discussed below.

The pronounced dependence of the NDR on the Sr doping level has been discussed in detail in [37]. In lightly doped LSCO, the reduction of the average Sr concentration is accompanied by an enhanced compositional and electronic inhomogeneity, leading to a strongly nonuniform spatial distribution of the electric field. As a result, regions with locally enhanced electric fields emerge, giving rise to local overheating of charge carriers and triggering the transition to the NDR regime at lower currents and higher temperatures in more weakly doped samples. Importantly, the observed NDR cannot be accounted for by standard field-induced effects within variable-range hopping and reflects a genuinely nonequilibrium transport response of a strongly disordered, percolative electronic system.

In oxygen-doped LCO+δ single crystals, Belevtsev *et al.* first observed strong nonlinearities in the current–voltage characteristics at temperatures $T \lesssim 8$–10 K, including the appearance of *S*-shaped *I–V* curves with regions of negative differential resistance once the applied current exceeds a threshold value $J_{th}$ [13]. In the low-bias (Ohmic) regime, the transport is well described by Mott variable-range hopping (VRH). Upon increasing the bias, however, the differential resistance drops sharply, signaling a breakdown of the VRH regime and the onset of nonlinear conduction [13] [35]. These effects were interpreted in terms of intrinsic electronic inhomogeneity associated with oxygen nonstoichiometry, leading to phase separation into regions of insulating and superconducting phases. In this picture, increasing the current can activate additional percolative conduction paths through superconducting inclusions embedded in an insulating matrix, resulting in a runaway reduction of voltage with increasing current. [13].

Subsequent studies demonstrated that qualitatively similar NDR phenomena can also occur in lightly Sr-doped LSCO, even though the microscopic origin is different. In particular, Dalakova *et al.* investigated LSCO crystals with $x \lesssim 0.01$ in the antiferromagnetic state and reported pronounced current-controlled NDR at low temperatures [35]. In contrast to LCO+δ, the authors emphasized that macroscopic phase separation with the formation of superconducting droplets is unlikely in Sr-doped samples due to the immobility of Sr ions and the absence of oxygen-driven chemical segregation [35]. Instead, the nonlinear transport and NDR were interpreted within a framework combining percolative hopping conduction, electron overheating, and the formation of localized "hot spots" in the vicinity of Sr-induced disorder. Local enhancement of the electric field and carrier temperature leads to a strong nonlinear increase of conductivity and the emergence of NDR above characteristic threshold fields of order $E_c \sim 5$–10 V/cm [35].

Magnetotransport measurements further support the distinction between these regimes. In lightly doped LSCO, Dalakova *et al.* observed a positive magnetoresistance at low bias currents and temperatures $T < 10$ K, which they associated with the influence of the antiferromagnetic and spin-density-wave background on hopping transport [35]. As the bias current or temperature is increased and the system enters the nonlinear/NDR regime, the magnitude of the positive magnetoresistance is reduced and can eventually change sign, reflecting the growing importance of nonequilibrium and percolative transport processes [35]. In oxygen-doped LCO+δ, the magnetic-field response in the nonlinear regime is additionally influenced by the presence of superconducting inclusions, leading to a more complex interplay between localization, superconductivity, and magnetic field effects [13] [32].

Overall, the occurrence of nonlinear transport and NDR in lightly doped La-based cuprates highlights the highly inhomogeneous and non-equilibrium nature of charge transport in the low-doping regime. While both oxygen-doped and Sr-doped systems exhibit qualitatively similar NDR phenomena, the underlying mechanisms differ substantially: superconducting phase separation plays a central role in LCO+δ, whereas in LSCO the effects can be consistently explained without invoking superconducting inclusions at these extremely low doping levels, where transport takes place in a predominantly antiferromagnetic background driven far from equilibrium by the applied electric field [13] [35].

For clarity, we emphasize that similar NDR features do not imply the same microscopic physics: in oxygen-rich $La_2CuO_{4+\delta}$, the nonlinearity is naturally linked to deep chemical phase separation and activation of percolative paths through superconducting inclusions, whereas in lightly Sr-doped $La_{2-x}Sr_xCuO_4$, the nonlinearity is governed by nonequilibrium hopping, electron overheating, and current-driven hot-spot formation in a strongly disordered antiferromagnetic background. It should also be noted that repeated I–V sweeps at fixed temperature yielded identical thresholds within experimental uncertainty, indicating negligible oxygen redistribution under the low-temperature measurement conditions.

### Sr *vs*. O-Doping – Disorder Nature and Phase Separation Depth

It is instructive to contrast Sr substitution ($La_{2-x}Sr_xCuO_4$) with excess oxygen doping ($La_2CuO_{4+\delta}$) in terms of the nature of disorder and the resulting superconducting landscape. Strontium substitution replaces $La^{3+}$ with $Sr^{2+}$ ions in the $La_2O_2$ layers, introducing quenched ionic disorder and a static Coulomb potential that couples to the $CuO_2$ planes. Because Sr ions are essentially immobile at experimental temperatures, the disorder they introduce is frozen-in and does not anneal upon cooling. To first approximation, the holes donated by Sr are distributed throughout the $CuO_2$ planes, so LSCO is often treated as a chemically homogeneous alloy; however, especially at low Sr concentrations, spatial fluctuations in the local Sr content and hole density are unavoidable, leading to nanoscale electronic and magnetic inhomogeneity. By contrast, interstitial oxygen atoms are mobile, can cluster and order, and locally distort the lattice, so oxygen-doped $La_2CuO_4$ tends to exhibit true chemical phase separation into oxygen-rich and oxygen-poor regions. The disorder introduced by Sr substitution is more quenched and random: at suboptimal doping levels, it is established during crystal growth, possibly influenced by electronic correlations accompanying bond formation, and subsequently remains frozen on experimental time scales. By contrast, the disorder associated with excess oxygen is annealable and correlated owing to the high mobility of interstitial oxygen atoms [47]. As a result, oxygen-related disorder can be modified by thermal treatment and even by long-term storage, giving rise to ordered superstructures or 'staging' patterns under appropriate conditions [35], or, conversely, to a gradual loss of excess oxygen, as evidenced by an increase of the Néel temperature. The depth of phase separation is therefore much greater in oxygen-doped samples: instead of a single disordered phase, two distinct phases with different lattice parameters and electronic properties can coexist [3] [44]. This fundamental difference provides a natural explanation for why oxygen doping can induce superconductivity at far lower average hole concentrations than Sr substitution. For example, even though the average hole concentration remains very low, an excess oxygen content as small as $\delta \approx 0.01$ can be sufficient to trigger a superconducting onset near 40 K in $La_2CuO_{4+\delta}$–based systems [3] [44], whereas Sr-doped LSCO with an equivalent hole concentration ($x \approx 0.03$) remains an antiferromagnetic insulator with no bulk superconductivity above a few kelvin [7] [9]. In oxygen-doped materials, the added oxygen tends to form oxygen-rich regions with locally high hole concentration, approaching optimal doping, while Sr-doped LSCO does not exhibit macroscopic chemical phase separation on experimental time scales [8]. As a result, superconductivity emerges gradually with increasing $x$ in Sr-doped LSCO, but appears abruptly at near-optimal $T_c$ once a percolating oxygen-rich phase is established in LCO+$\delta$. This distinction naturally follows from the different spatial scales and thermal coupling of inhomogeneities: in oxygen-doped LCO+$\delta$, mesoscale superconducting inclusions are weakly coupled and can be destabilized directly by applied current, whereas in Sr-doped LSCO the inhomogeneity is predominantly electronic and nanoscale, favoring current-induced local overheating (hot-spot formation) rather than direct suppression of superconducting regions.

From a granularity perspective, Sr- and O-doping also differ. Sr doping creates random nanoscale fluctuations in the hole density – the system may form charge stripe domains or patchy magnetism at very low $x$, but it does not typically break up into macroscopic superconducting and insulating regions at the same temperature [4]. In O-doped LCO, we effectively have a granular superconductor embedded in an insulator: the O-rich regions behave like SC "grains" with well-defined $T_c \approx 30\text{–}40$ K, weakly coupled through Josephson links across insulating boundaries [5] [13]. As a result, superconductivity in LCO+δ is inherently percolative and highly sensitive to external perturbations, in contrast to LSCO, where superconductivity emerges gradually with increasing $x$ and is controlled by disorder rather than chemical phase separation. This distinction is directly reflected in transport measurements: in oxygen-doped $La_2CuO_{4+\delta}$, current–voltage characteristics exhibit pronounced current-driven nonlinearity and negative differential resistance at low temperatures (Fig. 5), consistent with a percolative network of superconducting regions driven far from equilibrium.

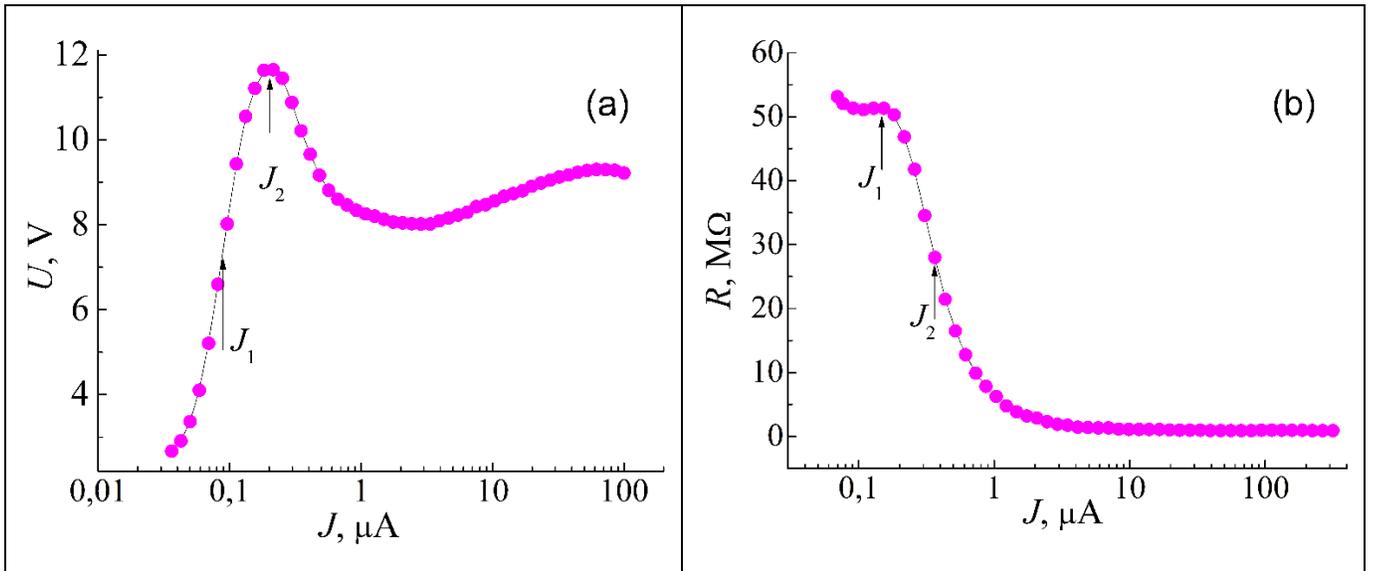

Fig. 5. Current-driven nonlinear transport and negative differential resistance in an oxygen-doped $La_2CuO_{4+\delta}$ single crystal at $T = 5.2$ K (data replotted from Ref. [32]). (a) Current–voltage characteristic exhibiting a pronounced $S$-shaped nonlinearity with two characteristic threshold currents $J_1$ and $J_2$ (inset: semilogarithmic scale). (b) Corresponding current dependence of the resistance, demonstrating an abrupt collapse of resistance upon entering the nonlinear regime.

Additional evidence for the percolative and nonequilibrium nature of nonlinear transport in oxygen-doped $La_2CuO_{4+\delta}$ is provided by magnetotransport measurements, which reveal a strong current-dependent and anisotropic magnetoresistance in the nonlinear regime (Fig. 6).

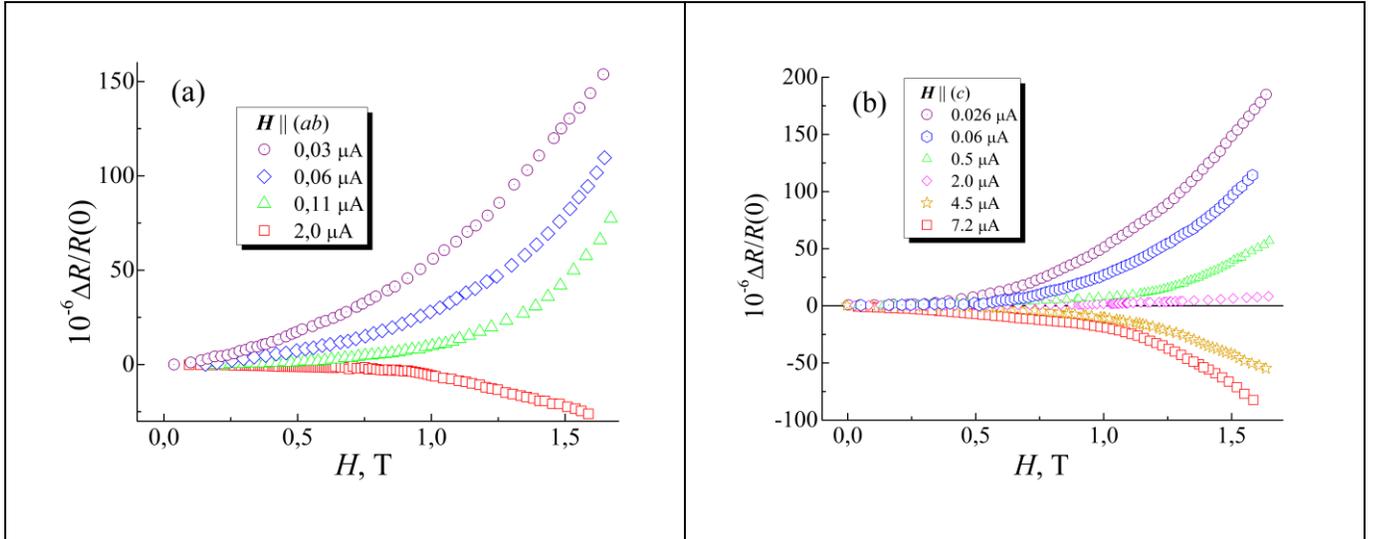

Fig. 6. Magnetoresistance of an oxygen-doped $La_2CuO_{4+\delta}$ single crystal at $T = 5$ K for different transport currents $J$ and magnetic-field orientations. (a) $H\|(ab)$; (b) $H\|c$. The strong current dependence and the sign change of the magnetoresistance correlate with the onset of nonlinear transport and negative differential resistance (data replotted from Ref. [32]).

This contrast provides a natural framework for understanding the qualitative differences in transport and magnetic response between Sr- and O-doped LSCO discussed below.

### Underdoped Superconducting Regime ($0.05 < x < 0.10$)

As the Sr concentration exceeds the critical threshold $x_c \approx 0.05$ [9], the system enters a regime where global phase coherence is established, yet the superconducting state remains deeply influenced by underlying magnetic and structural disorder. Transport measurements in this range ($0.05 < x < 0.10$) reveal a transition from insulating VRH behavior to a metallic-like normal-state transport [48].

A hallmark of this regime, inherited from the granular and percolative nature of superconductivity discussed above, is the persistence of the two-stage superconducting transition [7] in the resistivity $R(T)$. At $T_c^{onset}$, a sharp drop in resistivity signals the emergence of superconductivity within hole-rich clusters. However, zero resistance ($T_{c0}$) is achieved only at much lower temperatures, once Josephson coupling [22] bridges the gaps between these clusters. For samples near the percolation limit ($x \approx 0.06$), the transition width $\Delta T$ can span 10–15 K, reflecting the extreme spatial heterogeneity of the superconducting order parameter inferred from percolative scenarios of superconductivity [49].

Magnetic probes provide crucial insight into the nature of this inhomogeneity. µSR studies have demonstrated microscopic coexistence of spin-glass order and superconductivity in underdoped LSCO [8], while complementary NQR measurements reveal pronounced spatial variations of the local hole concentration, indicating strong electronic inhomogeneity on mesoscopic length scales [14]. For $x \approx 0.06$, a spin-freezing transition is observed at $T_g \approx 5$–8 K, occurring nearly simultaneously with the establishment of global superconducting coherence. This suggests a "patchy" landscape where superconducting puddles are interspersed with regions of frozen Cu spins. In this "cluster superconducting" state, the supercurrent must navigate a network of weak links sensitive to both thermal fluctuations and external magnetic fields.

Consequently, the application of a magnetic field leads to a dramatic broadening of the resistive transition. Unlike optimally doped samples, where the field primarily shifts $T_c$, in the underdoped regime, even modest fields ($H < 1$ T) can suppress the inter-cluster coupling [48], causing the $R(T)$ curves to exhibit a characteristic 'fan-like' broadening. At higher magnetic fields, superconductivity can be fully suppressed, revealing the intrinsically insulating normal state of underdoped LSCO [50]. This behavior highlights that in the underdoped regime, the global superconducting state is limited not by the pairing energy within the clusters, but by the fragility of the percolative network. Consistent with this picture, transport measurements show that increasing bias current or magnetic field readily suppresses the weakest inter-cluster links, further underscoring the low phase stiffness of the underdoped state.

Additional support for this picture comes from nonlinear transport studies by Belevtsev and Dalakova, who showed that in lightly doped cuprates the superconducting state can be destabilized by remarkably small bias currents, reflecting the breakdown of weak inter-cluster links rather than suppression of local pairing [13] [35]. Such current-driven effects provide an independent dynamical probe of the low phase stiffness and percolative nature of superconductivity in the underdoped regime.

### Near-Optimal Doping ($x \approx 0.15$)

Near optimal doping, La$_{2-x}$Sr$_x$CuO$_4$ enters the canonical *strange-metal* regime in its normal state, while the superconducting phase becomes far more coherent and homogeneous than in the underdoped case. Transport measurements show that the in-plane resistivity in this regime is approximately linear in temperature over a wide range. In our $x \approx 0.15$ samples ($T_c \approx 38$–40 K), $\rho(T)$ above $T_c$ follows an almost linear-in-$T$ dependence from just above the transition up to room temperature, consistent with earlier reports on optimally doped LSCO [19]. This behavior contrasts with that of underdoped samples (e.g., $x = 0.10$), where deviations from linearity or a weak upturn in $\rho(T)$ may appear upon cooling toward $T_c$, reflecting the influence of pseudogap physics (which sets in at much higher temperatures, $T^*$) or localization effects [9] [19].

The linear resistivity at optimal doping is widely discussed as a manifestation of non-Fermi-liquid transport, often interpreted in terms of Planckian dissipation or proximity to a putative quantum critical point [19] [51]. What is clear experimentally is that conventional Fermi-liquid behavior ($\rho \propto T^2$) is not observed at $x \approx 0.15$ over the accessible temperature range; instead, the scattering rate scales approximately linearly with temperature, reflecting the anomalous nature of the strange-metal state [19] [51]. At the same time, spectroscopic probes indicate that the pseudogap is strongly weakened near optimal doping. Complementary NQR and related local-probe measurements indicate pronounced nanoscale spatial variations in the local electronic environment in Sr-doped LSCO, with a tendency toward greater homogeneity as optimal doping is approached [14].

Despite this increased homogeneity, superconducting fluctuations persist above $T_c$ even at optimal doping. Nernst-effect and torque-magnetometry experiments on LSCO and related cuprates demonstrate that a precursor diamagnetic response survives up to ~10–15 K above $T_c$, reflecting short-lived superconducting correlations in the normal state. Such behavior is naturally interpreted in terms of phase-disordered superconductivity, in which pairing amplitude remains finite while long-range phase coherence is lost at $T_c$, as directly demonstrated by high-frequency conductivity measurements in underdoped cuprates by Corson et al. [23]. This fluctuation regime is significantly narrower than in strongly underdoped LSCO, where phase fluctuations extend over a much broader temperature interval, but it underscores that superconductivity at $x \approx 0.15$ is still not fully described by a simple mean-field BCS picture.

In contrast to the underdoped regime, however, these fluctuations do not impede the formation of a robust global superconducting state. Transport and magnetization measurements at $x \approx 0.15$ reveal a sharp superconducting transition: the resistive transition width (difference between onset and zero resistance) is typically on the order of a few kelvins or less, with no evidence for a distinct second drop in $\rho(T)$ [7] [9].

Likewise, AC susceptibility and magnetization exhibit a single, well-defined diamagnetic transition at $T_c$, without the long tails or multiple features characteristic of granular or percolative superconductivity [7]. This behavior indicates that the superconducting state at optimal doping is spatially homogeneous on macroscopic length scales and characterized by a much higher phase stiffness than in the underdoped regime.

The reduced role of disorder and granularity at optimal doping is further reflected in the weak sensitivity of the superconducting transition to external perturbations. Applying modest magnetic fields (of order 0.1 T) or varying the measurement current has only a minor effect on $T_c$ or the transition width in $x \approx 0.15$ samples, in sharp contrast to lightly doped crystals where similar perturbations dramatically suppress superconductivity [48]. Consistently, we do not observe pronounced current-induced nonlinearities or negative differential resistance in the $I$–$V$ characteristics of the optimally doped samples, whereas such effects are readily detected at lower Sr concentrations and have been associated with the breakdown of weak inter-cluster links in a granular superconducting network [13] [35].

A direct comparison between slightly underdoped ($x = 0.10$) and optimally doped ($x = 0.15$) LSCO highlights this evolution from a disorder-dominated to a homogeneous superconducting state. The $x = 0.10$ sample ($T_c \approx 30$ K) still exhibits a broadened transition [9], with a resistive onset well above the temperature where zero resistance is achieved, and shows residual signatures of pseudogap physics in transport, together with local-probe (NQR) evidence for persistent electronic inhomogeneity at this doping level [14]. By contrast, at $x \approx 0.15$ the onset of pairing and the establishment of global phase coherence nearly coincide, and the normal-state transport displays no discernible pseudogap-related anomalies. This trend is consistent with a wide body of experimental work indicating that superconducting connectivity in LSCO becomes essentially complete by $p \sim 0.12$–$0.15$, as the last remnants of the antiferromagnetic insulating phase disappear [7].

In summary, near the optimal doping level, the effects of disorder and granularity are significantly reduced, yielding a superconducting state that is robust and spatially homogeneous, as well as a normal state that exhibits the enigmatic strange-metal behavior. This regime marks the crossover from percolative, weakly phase-stiff superconductivity at low doping to a coherent superconducting condensate embedded in a single-phase metallic background.

### Overdoped Regime ($x > 0.18$)

For completeness, we briefly summarize the key phenomenological features of the overdoped regime in $La_{2-x}Sr_xCuO_4$, which serves as a reference limit where disorder- and granularity-driven effects largely disappear. On the overdoped side of the phase diagram, LSCO gradually evolves toward a more metallic state in the normal phase, while the superconducting state becomes spatially homogeneous and mean-field-like. Transport measurements indicate that, for $x \gtrsim 0.18$, the low-temperature normal-state resistivity approaches a Fermi-liquid form, $\rho(T) \propto T^2$, as inferred from the behavior just above $T_c$ and from systematic transport studies in the overdoped regime [19]. When superconductivity is fully suppressed by strong magnetic fields, the resulting normal state is metallic, consistent with a Fermi-liquid-like ground state [50]. This behavior contrasts sharply with the linear-in-$T$ resistivity observed near optimal doping, indicating that electron–electron scattering in a coherent metallic state dominates transport. At the same time, the pseudogap is no longer observed in this regime: spectroscopic and thermodynamic probes show that by $p \approx 0.18$–$0.20$ the pseudogap temperature $T^*$ merges with $T_c$ and disappears beyond the superconducting dome [19]. Angle-resolved photoemission and transport measurements further indicate that the large, hole-like Fermi surface characteristic of a conventional metal is fully restored in overdoped LSCO [10].

The superconducting properties in the overdoped regime likewise reflect this increased homogeneity. Although $T_c$ decreases with increasing Sr content beyond the optimal doping level, following the familiar dome-shaped phase diagram, the superconducting transition remains sharp and single-step even at high Sr concentrations. Transport and magnetic measurements reveal narrow resistive and susceptibility transitions, without the broad tails, two-stage behavior, or anomalous features characteristic of the underdoped granular regime. Nonlinear transport phenomena, such as current-induced instabilities or negative differential resistance observed at low doping, are absent: the *I–V* characteristics are well described by conventional ohmic behavior in the normal state and by a well-defined critical current in the superconducting state. Likewise, moderate magnetic fields suppress superconductivity in a manner consistent with standard vortex physics, without revealing signatures of weak-link networks or competing magnetic order, in sharp contrast to the strongly inhomogeneous underdoped regime.

Overall, the overdoped regime of LSCO represents a transition to a relatively simple, single-phase metallic superconductor, in which disorder plays only a minor residual role due to strong electronic screening at high carrier densities. This regime therefore provides a useful reference point for the preceding discussion: as the Sr concentration is reduced from overdoped to underdoped compositions, the system evolves continuously from a homogeneous Fermi-liquid-like metal into a fragmented, disorder-dominated superconductor characterized by electronic and chemical phase separation, granularity, and percolative transport. In the following section, we build on this experimental phenomenology to discuss theoretical frameworks capable of capturing this evolution, with particular emphasis on the roles of disorder, localization, and interaction effects across the phase diagram.

### Unified phenomenology across the LSCO phase diagram

Although the preceding subsections considered different doping regimes of $La_{2-x}Sr_xCuO_4$ separately, the overall experimental phenomenology across the phase diagram exhibits a remarkable internal coherence. Transport anomalies observed at low doping, granular superconductivity in the underdoped regime, and the eventual recovery of a homogeneous superconducting state at higher carrier concentrations should not be viewed as unrelated or competing phenomena. Rather, they represent successive manifestations of a common underlying tendency toward spatially inhomogeneous electronic states in the presence of disorder, in which percolative transport can mask the underlying microscopic mechanisms and mimic canonical transport laws, as demonstrated in other strongly disordered and granular systems, including magnetic oxide composites [52] [53].

At the lowest doping levels, this tendency is revealed through strong carrier localization and variable-range hopping transport, indicating that the electronic system has already broken up into regions with markedly different local conductivities. Upon increasing hole concentration, superconductivity does not emerge uniformly but instead nucleates locally within hole-rich regions, giving rise to granular superconductivity and percolative charge transport. The characteristic two-stage superconducting transition, broad resistive tails, nonlinear current response, and pronounced sensitivity to weak magnetic fields all point to a superconducting state whose global coherence is limited by the connectivity between locally superconducting regions rather than by the strength of pairing itself.

With further doping, the same system gradually evolves toward a coherent superconducting state in which spatial inhomogeneity becomes less relevant for macroscopic transport. The disappearance of two-step transitions, nonlinear transport anomalies, and history-dependent magnetoresistance signals marks the crossover from a percolative, weakly phase-stiff superconducting network to a homogeneous superconductor with robust global phase coherence. Importantly, this evolution does not rely on the assumption of distinct superconducting mechanisms in different doping regimes; rather, it can be viewed as a continuous transformation driven by the progressive weakening of localization and disorder-induced granularity.

From this perspective, disorder plays a unifying role across the entire LSCO phase diagram. Rather than acting merely as a perturbation superimposed on an otherwise uniform electronic system, disorder actively shapes the electronic ground state by promoting spatial variations in carrier density, superconducting stiffness, and magnetic correlations. The balance between localization, granularity, and coherence is then controlled primarily by doping, which tunes the effectiveness of electronic screening and the strength of inter-region coupling. This framework provides a natural phenomenological link between the insulating, underdoped, optimally doped, and overdoped regimes discussed above and sets the stage for a more detailed discussion of the specific roles played by different types of disorder and transport nonlinearity in the following sections.

## Discussion

The experimental trends observed across lightly doped, underdoped, and near-optimally doped LSCO can be consistently discussed within a broad phenomenological framework that emphasizes the interplay between disorder, electronic correlations, and spatial inhomogeneity. From this perspective, the diverse transport regimes observed in LSCO (from a Mott-like insulating state at the lowest hole concentrations, through granular and percolative superconductivity near the threshold of bulk superconductivity, to the strange-metal regime close to optimal doping) are best viewed as continuously evolving crossover regimes. Their evolution reflects the combined effects of doping-dependent screening, disorder, and inter-region coupling, rather than a sequence of distinct phases governed by a single microscopic mechanism [7].

To make this continuity more explicit, it is useful to frame the crossover in LSCO as a multiscale process linking microscopic scattering to mesoscale inhomogeneity and macroscopic connectivity. At a conceptual level, the experimental crossover from hopping-dominated transport at low doping to strange-metal behavior near optimal doping can be viewed as a multiscale process. On the microscopic level, strong (near-unitary) scattering from dopant-related disorder generates resonant states and strongly energy-dependent quasiparticle lifetimes [54]. At low carrier density, poor electronic screening amplifies the spatial inhomogeneity induced by such scattering, resulting in electronically granular landscapes where transport proceeds via hopping and weak inter-region links. As doping increases, improved screening progressively suppresses long-range Coulomb disorder and enhances inter-region coupling. From a macroscopic perspective, this evolution can be interpreted as a percolative crossover, in which spatially coherent conduction paths gradually emerge and eventually dominate transport. Within this view, the emergence of $T$-linear resistivity does not require a change in the underlying scattering mechanism, but rather reflects the formation of spatially coherent transport pathways governed by strongly scattered quasiparticles.

At the lowest dopings, poor electronic screening and strong scattering centers (such as $Sr^{2+}$ substitutional defects and oxygen-related disorder) confine charge carriers to localized states, giving rise to insulating behavior and variable-range hopping transport dominated by interaction and spin-disorder effects [9] [38] [43]. With increasing hole concentration, disorder and local fluctuations promote the formation of spatially segregated, hole-rich regions in which superconductivity nucleates locally, while global phase coherence remains limited by weak inter-cluster coupling—defining a granular superconductivity regime that can be naturally described within a percolative framework [49], with the superconducting transition governed by phase fluctuations and Josephson coupling between clusters [22]. Upon further doping toward optimal concentration, enhanced electronic screening and stronger coupling between superconducting regions progressively weaken localization and connectivity barriers, allowing the system to cross over into a coherent, spatially homogeneous superconducting state embedded in a metallic background [9]. Crucially, this entire evolution reflects a continuous crossover governed by disorder strength, screening efficiency, and coupling between regions, rather than by an abrupt change in a single microscopic scattering mechanism. In this sense, disorder and its screening-induced attenuation serve as a smooth interpolating parameter between the Mott-insulating and strange-metal limits of LSCO.

We stress that impurity- and disorder-based interpretations of LSCO transport should be regarded as phenomenological frameworks - useful interpretive lenses rather than uniquely established or exhaustive theories. They provide coherent ways to connect microscopic disorder and scattering processes with macroscopic transport behavior, but they do not exclude alternative or complementary interpretations. Indeed, several influential frameworks have been advanced to explain key aspects of LSCO's transport. The linear-in-temperature resistivity near optimal doping is often discussed in terms of Planckian dissipation, where the scattering rate approaches a putative quantum limit of order $\hbar/(k_B \cdot T)$ [51], and the strange-metal state is viewed as intrinsically quantum-critical [19]. Closely related are quantum-critical scenarios invoking a hidden zero-temperature phase transition, such as the termination of the pseudogap phase at a critical doping, as the origin of non-Fermi-liquid transport [19]. In addition, phase competition and intertwined orders provide another important perspective: in underdoped cuprates, charge-stripe correlations and related spin-ordering phenomena can coexist with or compete against superconductivity [4]; in oxygen-doped La-based cuprates, static spin-density-wave order has been shown to emerge concurrently with superconductivity [5]. Local probes reveal microscopic phase complexity at low doping levels: μSR demonstrates coexistence of magnetism and superconductivity [8], while NQR highlights strong spatial inhomogeneity of the local electronic environment [14].

It is also important to separate the near-equilibrium, linear-response strange-metal regime (where Planckian arguments are typically formulated in the context of linear-response transport) from the high-field, far-from-equilibrium NDR regime discussed above. The latter corresponds to a strongly nonequilibrium transport regime with local electron overheating and hot-spot formation and is fundamentally different from the low-field transport behavior inferred from $\rho(T)$.

The disorder-centric approach adopted here offers a complementary viewpoint that naturally incorporates many of these phenomena without invoking fine-tuned criticality in an ideally clean system. Rather than attributing anomalous transport solely to universal quantum limits or intrinsic instabilities, it emphasizes that real-world disorder, unavoidable in LSCO, can generate pronounced nanoscale electronic inhomogeneity, as directly evidenced by $^{63}$Cu NQR measurements [14]. Such spatial variability provides a natural setting for percolative transport and connectivity-driven crossovers, particularly in oxygen-doped and phase-separated compositions [5], and it also creates a heterogeneous background in which magnetic and charge correlations can develop and interact with superconductivity [4] [8]. From this perspective, linear-in-$T$ ("Planckian-like") resistivity may still reflect a maximal effective scattering rate [51], while its microscopic origin can remain multifactorial and sensitive to disorder, screening, and mesoscale connectivity. Accounting explicitly for spatial heterogeneity introduced by both quenched and annealable disorder, therefore, helps explain why the evolution from insulating to metallic and superconducting behavior in LSCO is gradual and domain-structured, rather than a sequence of sharp phase transitions predicted by idealized models.

A central outcome of this discussion is the recognition that disorder, both static disorder from Sr substitution and dynamic, correlated disorder from oxygen intercalation, plays a unifying role across the LSCO phase diagram. Sr substitution introduces a frozen electrostatic potential and local lattice strain, while excess oxygen creates mobile, cluster-forming defects that promote mesoscale chemical phase separation [3] [44] [47]. Both routes lead to pronounced spatial variations in the electronic state on mesoscopic length scales: local variations of the hole concentration are directly evidenced by NQR measurements [14], while scanning probe studies reveal strongly inhomogeneous diamagnetic and superconducting responses consistent with an underlying electronic phase separation [55]. These variations naturally couple magnetic, superconducting, and transport properties. In lightly doped LSCO, the combination of disorder and antiferromagnetic correlations localizes carriers, leading to insulating transport that is strongly tied to magnetic order [7] [9] [11]. Near the superconducting threshold, the same inhomogeneity yields a mixed landscape of magnetic regions and superconducting clusters. Global superconductivity then emerges through a percolative connectivity of these clusters, as expected for an intrinsically inhomogeneous superconducting state [49], while the establishment of long-range phase coherence, and hence zero resistance, is governed by weak-link (Josephson) coupling between them [15]. The resulting superconducting transition is therefore broad, often two-stage, and highly sensitive to current and magnetic field, reflecting the prominent role of phase fluctuations in a system with low superfluid stiffness [22]. By optimal doping, improved screening suppresses large-scale inhomogeneity, yet residual disorder continues to shape the strange-metal transport through anomalous scattering mechanisms discussed in the broader literature [19]. Disorder thus acts not as a passive background but as an active ingredient that continuously reshapes the electronic state.

It is noteworthy that even in strongly disordered and granular regimes, hallmark fluctuation phenomena of cuprate superconductivity persist. In highly inhomogeneous LSCO samples, transport and thermodynamic probes reveal signatures of superconducting fluctuations above the bulk $T_c$, including paraconductivity and a finite Nernst response, indicating the presence of short-lived Cooper pairing and vortex-like excitations well before global phase coherence is established [21]. Independently, high-frequency conductivity measurements in underdoped cuprates have shown that phase coherence is lost at $T_c$, while local pairing correlations persist to substantially higher temperatures, reflecting a regime of phase-incoherent superconductivity dominated by vortex dynamics [23]. Similar fluctuation effects are well documented in cleaner cuprate crystals, where Nernst and diamagnetic signals can extend up to $\approx 1.5 \cdot T_c$. Their qualitative persistence in disordered and granular samples underscores the robustness of local pairing correlations in the $CuO_2$ planes: disorder primarily modifies spatial coherence and percolative connectivity, but does not eliminate the underlying tendency toward Cooper pairing. Although the pseudogap and fluctuation regimes may be altered in extent or visibility by disorder and phase separation, they remain experimentally detectable, indicating that moderate disorder leaves the core superconducting mechanism intact at the local level [21].

By synthesizing extensive experimental results from LSCO single crystals and ceramics, including a substantial body of work obtained at the B. Verkin Institute for Low Temperature Physics and Engineering, and interpreting them within this disorder-aware framework, the present review clarifies how magnetism, superconductivity, and transport in LSCO are intertwined across doping [11] [35] [37]. Viewing superconductivity as emerging through percolation in an inhomogeneous medium helps reconcile two-stage superconducting transitions with a unified physical picture [15] [49], while providing a natural framework for interpreting nonlinear transport phenomena and negative differential resistance observed in strongly inhomogeneous LSCO [35]. At the same time, interpreting the strange-metal transport and the fading of the pseudogap near optimal doping as consequences of improved delocalization and screening provides a disorder-informed complement to quantum-critical and Planckian paradigms [19] [51]. Finally, this perspective suggests concrete experimental directions: systematic tuning and characterization of disorder (via controlled quenching, irradiation, or defect engineering) could directly test its role as a continuous control parameter that connects the insulating, granular, and strange-metal regimes. Overall, a nuanced consideration of disorder emerges as essential for a holistic understanding of LSCO and, more broadly, of superconductivity in strongly correlated and imperfectly ordered materials.

### Conclusion

In this review, we have traced the evolution of charge transport $La_{2-x}Sr_xCuO_{4+\delta}$ across experimentally accessible doping regimes, with a particular focus on how disorder and spatial inhomogeneity reshape the interplay between antiferromagnetism, superconductivity, and metallic conduction. The central empirical message is that LSCO should not be viewed as a sequence of sharply separated "textbook" phases realized in an idealized, homogeneous host. Over a substantial part of its phase diagram, especially on the underdoped side, the electronic system is intrinsically heterogeneous, and the macroscopic transport response is governed as much by connectivity, percolation, and nonequilibrium effects as by the local pairing scale or the average hole density.

At the lowest dopings, the normal state is dominated by strong localization and hopping transport, with resistivity behavior that is tightly correlated with the magnetic background. As doping increases, superconductivity emerges first locally (in spatially confined, hole-rich regions) well before global phase coherence is established. This naturally produces broad, two-stage resistive transitions, extreme sensitivity to current and magnetic field, and a wide parameter range in which superconducting and magnetic responses coexist. In oxygen-enriched samples, where mobile interstitial oxygen can self-organize and phase-separate, the resulting granularity is deeper and more mesoscale in character; in Sr-doped LSCO, the disorder is quenched, and the inhomogeneity is predominantly electronic, yet still sufficient to generate a "patchy" superconducting landscape near the percolation threshold. These observations underscore that the route to superconductivity in underdoped LSCO is often percolative and network-limited, rather than a uniform, mean-field-like condensation.

A defining contribution of the experimental body of work summarized here, much of it developed in the ILTPE (Kharkiv) program, is the identification and systematic characterization of nonlinear transport phenomena in the strongly inhomogeneous regime. Current-driven departures from Ohmic response, including negative differential resistance in oxygen-rich, lightly doped samples, provide a dynamical probe of the fragile connectivity of the conducting (and potentially superconducting) network. These nonlinearities are not merely technical curiosities: they reveal that the low-doping state can be poised near instability boundaries where modest bias reorganizes current pathways, activates percolative channels, or drives nonequilibrium delocalization in a spatially nonuniform medium. In this sense, nonlinear transport complements conventional linear-response measurements by exposing the "hidden" structure of the conducting network and its sensitivity to disorder, thermal history, and magnetic field.

Near optimal doping, LSCO enters the canonical strange-metal regime with approximately linear-in-$T$ resistivity and a much more coherent superconducting transition than in the underdoped case. While the normal state becomes more homogeneous on macroscopic scales, fluctuation phenomena remain observable above $T_c$, indicating that superconducting correlations can persist locally even when global coherence is lost. The persistence of such fluctuation signatures across samples with different degrees and types of disorder emphasizes an important point: disorder strongly controls connectivity and phase coherence, but it does not trivially eliminate the underlying correlation physics of the $CuO_2$ planes. Any successful description of LSCO must therefore reconcile two facts that are sometimes discussed separately: robust local correlation phenomena on the one hand, and pronounced disorder-driven granularity and percolation on the other.

To organize this phenomenology, we adopt a disorder-aware interpretive perspective, in which resonant nonmagnetic impurity scattering in the unitary limit provides a unifying framework for understanding the crossover from localization-dominated transport to coherent metallic and strange-metal behavior. We have been careful to treat this as a phenomenological lens rather than an exclusive or uniquely validated microscopic theory. Nonetheless, it provides a useful way to connect several experimentally salient features (energy-dependent quasiparticle coherence, doping-dependent screening, and the progressive collapse of granularity) within a single narrative. Importantly, positioning disorder as an active control parameter does not conflict with other major paradigms in the cuprate literature (Planckian dissipation, quantum criticality, and phase competition); rather, it highlights how these intrinsic many-body tendencies are expressed in real materials whose electronic landscape is shaped by static and dynamic disorder.

Several open questions remain, and they are experimentally addressable. A key challenge is to establish quantitative criteria for the crossover from granular/percolative superconductivity to effectively homogeneous superconductivity, and to identify which disorder metrics (dopant configuration, oxygen ordering kinetics, nanoscale hole-density variance, or weak-link topology) best predict that crossover. Likewise, it remains important to disentangle intrinsic nonequilibrium electronic mechanisms from self-heating in nonlinear transport, ideally through combined measurements of *I–V* characteristics, noise spectra, local thermometry, and spatially resolved probes. More broadly, the disorder-centric viewpoint motivates targeted experiments that tune and characterize disorder in a controlled manner (via quench protocols, irradiation, defect engineering, or designed oxygen-ordering procedures) to test whether disorder can indeed function as a continuous tuning parameter connecting insulating, granular, and strange-metal regimes.

In summary, LSCO offers a uniquely informative laboratory for investigating how disorder and strong correlations jointly influence transport and superconductivity in a doped Mott system. The experimental evidence reviewed here supports a picture in which spatial inhomogeneity and percolation are not incidental complications but central organizing principles across much of the underdoped phase diagram. Recognizing this offers a coherent way to interpret a wide range of linear and nonlinear transport phenomena, and it sharpens the experimental agenda for clarifying how high-$T_c$ superconductivity emerges from the complex, disorder-influenced landscape of the cuprates.

## Acknowledgments

The author gratefully acknowledges support from the SARU (Scholars at Risk Ukraine) program, funded by the Carlsberg Foundation, which made it possible to continue research activities under challenging circumstances. The author also thanks Prof. Juan Bartolomé and Prof. Valentyna Sirenko for initiating this review and for stimulating discussions, as well as Prof. H.-G. Rubahn for his continued interest and support. The author would also like to honor the memory of his late colleagues Boris Belevtsev and Nina Dalakova, whose contributions to the study of transport phenomena in complex oxides, including LSCO, played an important role in shaping the scientific direction of this work.